\documentclass[12pt]{article}
\usepackage{amssymb}
\usepackage{amsthm}
\usepackage[version=3]{mhchem}
\usepackage{graphicx,graphics}

\begin{document}

\title{\textbf{Drawing a Waddington landscape to capture dynamic epigenetics\footnote{Reference: Nicol-Benoit, F., le Goff P., Michel D. 2013. Drawing a Waddington landscape to capture dynamic epigenetics. Biol. Cell. 105, 576-584.}}}
\author{\textbf{Floriane Nicol-Benoit, Pascale le Goff and Denis Michel} \footnote{Universit\'e de Rennes 1 IRSET U1085 Transcription, Environment and Cancer, Campus de Beaulieu, Bat. 13, 35042, Rennes Cedex, France, denis.michel@live.fr
}}
\date{}
\maketitle

\textbf{Abstract}. Epigenetics is most often reduced to chromatin marking in the current literature, whereas this notion was initially defined in a more general context. This restricted view ignores that epigenetic memories are in fact more robustly ensured in living systems by steady state mechanisms with permanent molecule renewal. This misconception is likely to result from misleading intuitions and insufficient dialogues between traditional and quantitative biologists. To demystify dynamic epigenetics, its most famous image, a Waddington landscape and its attractors, are explicitly drawn. The simple example provided is sufficient to highlight the main requirements and characteristics of dynamic gene networks, underlying cellular differentiation, de-differentiation and trans-differentiation.\\
\newline
\textbf{Running title:} Draw me a landscape\\
\textbf{Key words:} epigenetics, steady state, multistability \\
\newpage

\section{Alternative conceptions of epigenetics}
The dialogue between traditional and theoretical biologists is often limited and sometimes confused by ambiguous vocabulary, such as for the term epigenetics. Specialists of chromatin (the complex that contains DNA in the nucleus of eukaryotic cells), ascribed in good faith this word to their domain, based on the fact that it contributes to specify the phenotype though not directly inscribed in DNA sequence. For instance, journals whose name refer to "epigenetics" deal exclusively with chromatin. However, reducing epigenetics to the field of chromatin poses serious problems: (\textbf{i}) changes in chromatin are not very strong memories; (\textbf{ii}) conversely, strong memories can be generated without any modification of chromatin; (\textbf{iii}) the concept of epigenetics has been defined long before studies on chromatin, as a way of stabilizing a cellular state through the formation of attractors in dynamic gene networks. In the present context, attractors should be understood as compatible configurations of gene expression, that is to say gene expression patterns that can remain stable in absence of strong perturbations. A prerequisite for understanding dynamic epigenetics is a good perception of the steady state, characteristic of living systems and which should not be confused with a dynamic equilibrium.

\section{Intuition is not always a good adviser}
Despite the existence of pedagogical reviews on attractors (Huang and Ingber, 2006; Ferrell, 2012), it remains difficult for the biologist community to admit that nonequilibrium structures can not only be stable, but also ensure long-term memory. Setting a flag or a label appears at first glance as the only way to guarantee a permanent marking, whereas dynamic figures like vortices seem inherently fleeting. This impression is reinforced by DNA, that is actually a digital memory, effectively duplicated, repaired and propagated. Given that the different cell types in our body contain the same DNA, it is tempting to extend this principle to chromatin, onto which cell type-specific memories are supposed to be printed. But experiences showed that changes of chromatin marks remain possible in highly compacted chromatin and are too labile and reversible over short time scales to ensure a persistent memory. In fact, since the discovery of enzymes of chromatin modification and demodification, there is no difference between histone modifications and, for example, the phosphorylation of cyclins, long established as reversible. Curiously, to regain some stability of the cell differentiation states, we must return to other principles of memory engraved in moving figures, less intuitive because they seem more fragile. Nevertheless, we must admit that the persistence of dynamic circuits is the definition of life, but its understanding requires a good vision of the concept of steady state.

\section{The steady state}
The essential property of living matter is to be maintained in nonequilibrium states called steady states. An example of nonequilibrium structure is the vortex that forms above the drain of a bathtub which is emptied. This structure can be maintained if the bathtub plug is open and the inlet water flow corresponds exactly to the exhaust flow. This is precisely what is a living cell: an open nonequilibrium system sustained by permanent replenishment and dissipation of matter and energy. In the cell, the long-term storage of information, such as the cell differentiation status, is more registered in eddies than in labels. Molecular labeling exists but paradoxically ensures only short-term memory. This inversion of time scales compared to our everyday experience may appear disconcerting, but a fundamental feature of life is that these components are replaced very quickly. Except for a few structures for which molecule renewal is "technically" difficult or impossible, as for our cristallin, most of our constituent molecules are renewed within some months without us having the impression of having changed. It must be admitted that life and our identity are "sustained transient figures", robust versions of the vortex of the emptying bath. If the bathtub faucet is open, the vortex may appear similar on photos taken at different times, while the water molecules materializing it are constantly renewed. This continuous flow is the essence of life at all scales: at the cellular level, mRNA and proteins which are continuously degraded and resynthesized, and at the species level, individuals die and others are born. The only long-term characteristic of life is basically information (Michel, 2013). Matter comes alive in this information or more exactly, this information embodies in matter.

\section{Common misconceptions}
Several misconceptions must be ruled out (\textbf{i}) The steady state resembles, but is not an equilibrium, because it vanishes as soon as the system is disconnected from its environment, as for the vortex described above. But everyone knows that an active cell is never disconnected, except accidentally and definitely. (\textbf{ii}) At the microscopic level, the steady state is not more dynamic than a dynamic equilibrium. A bucket of water is a system in equilibrium where water molecules constantly move but can never spontaneously form a macroscopic structure like a whirlpool. (\textbf{iii}) Finally, a major feature of the steady state compared to equilibrium, is the possibility of getting different figures starting from the same ingredients, provided that some conditions are satisfied, such as the presence of feedback circuitry. Contrary to the steady state, a single final state is possible in equilibrium, irrespective of the number and nature of the ingredients involved. This phenomenon, essential for understanding systems biology and cellular differentiation, is called multistability. Waddington's epigenetics (Slack, 2002) is the most striking image of multistability, where different cell types can be obtained from the same batch of genes and without need for labeling them. Nonequilibrium systems have amazing and interesting properties. They can be very sensitive to disturbances that can cause sudden phenomena, asymmetries, special structures, in short, organization. The arrow of time and life appear out of equilibrium and as stated by the Nobel prize recipient Ilya Prigogine: "Out of equilibrium, matter begins to see".

\section{Building an elementary Waddington "epigenetic landscape"}
The best way to demystify Waddington landscapes is to build one explicitly, using simple mathematical tools. For this, let us choose an elementary network of genes, in fact the simplest possible one, made of a single gene encoding a transcription factor $ F $ and a single circuit, precisely a positive feedback on its own gene (Fig.1). The formal description of this prototype of positive loop, first proposed in (Keller, 1995), is modified below to be expressed as a function of the total concentration of $ F $. It is assumed that this factor binds to DNA only as a dimer, which is often the case for transcription factors.

\begin{center}
\includegraphics[width=6cm]{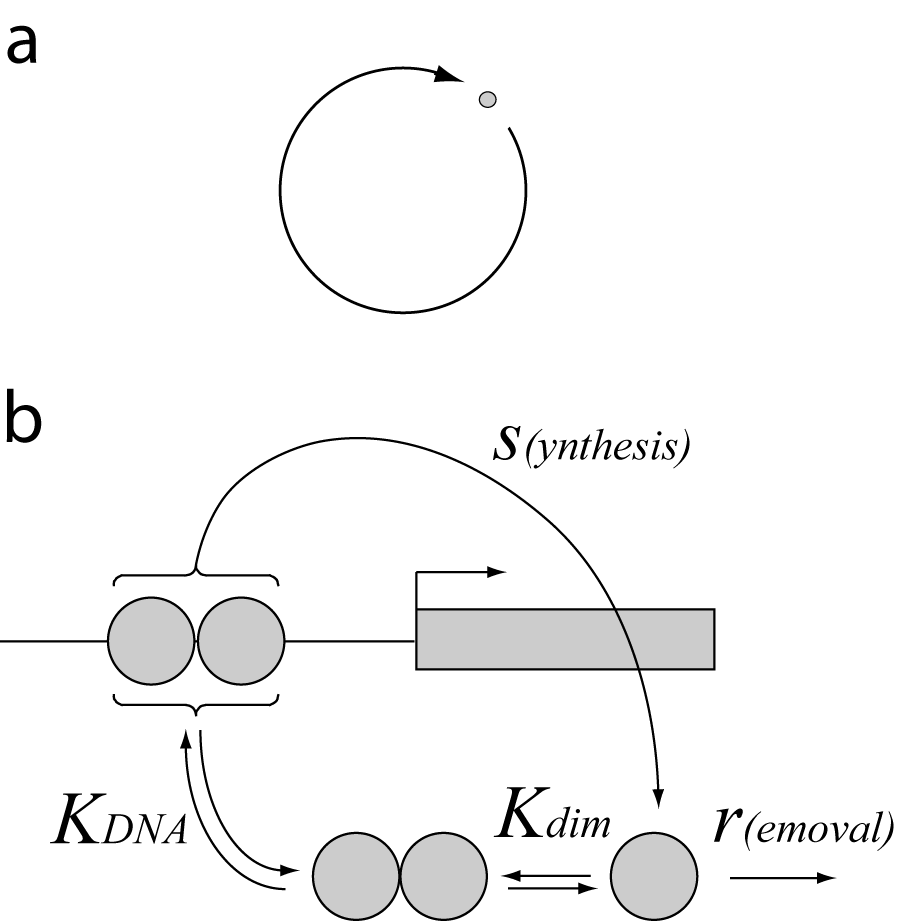} \\
\end{center}
\begin{small} \textbf{Figure 1. Schematic circuit used for the present model.} \\
(\textbf{a}) Mini network reduced to a single gene and a single circuit. (\textbf{b}) Detailed circuit including the parameters necessary for its modelling. A transcription factor is capable, as a dimer, of stimulating the transcription of its own gene. $ K_{DNA} $ is the equilibrium constant of dissociation from DNA and $ K_{dim} $ is the dimerisation constant. In the equations, these constants are written $ K $ and $ D $ respectively. \end{small}

\subsection{Quasi-equilibrium of DNA binding}
If the interactions between the transcription factor and the DNA are fast enough compared to gene expression dynamics, then the approximation of "time scale separation" is allowed (Michel, 2009) and the equilibrium constant can be assimilated to a concentration ratio. The dimensionless dissociation tendency depends on the absolute equilibrium dissociation constant $ K $ and on the concentration of $ [F_{2}] $ and corresponds to the ratio of free over bound DNA as follows

\begin{equation}   \frac{K}{[F_{2}]}=\frac{[DNA_{0}]}{[DNA-F_{2}]} \end{equation} 

where $ [DNA_{0}] $ and $ [DNA-F_{2}] $ should be understood as the times during which the promoter is free or occupied respectively.

DNA can be either empty or occupied by a dimer but not a monomer of $ F $. Hence, $ [DNA]_{tot}=[DNA_{0}]+[DNA-F_{2}] $ and its fraction of occupancy $ Y $ is 
the fraction of time during which DNA is occupied by $ F_{2} $

\begin{subequations} \label{E:gp}
\begin{equation}  Y=\frac{[DNA-F_{2}]}{[DNA_{0}]+[DNA-F_{2}]} \end{equation}  \label{E:gp1}
that can be transfomed using the value of $ [DNA-F_{2}] $ derived from Eq.(1), into
\begin{equation}   Y=\frac{[DNA_{0}][F_{2}]}{K[DNA_{0}]+[DNA_{0}][F_{2}]} \end{equation} \label{E:gp2}
which simplifies to
\begin{equation}   Y=\frac{[F_{2}]}{K+[F_{2}]} \end{equation} \label{E:gp3}
\end{subequations} 

\subsection{Quasi-equilibrium of dimerisation}
The dimerisation constant $ D $ can be expressed as
\begin{subequations} \label{E:gp}
\begin{equation}   D=\frac{[F_{2}]}{[F_{1}]^{2}} \end{equation} \label{E:gp1}
and the sum of monomeric and dimeric receptor is the total receptor
\begin{equation} [F]_{\textup{tot}}=2[F_{2}]+[F_{1}] \end{equation} \label{E:gp2}
\end{subequations} 

Eqs(3a,3b) give
\begin{equation} [F_{2}]=D([F]_{\textup{tot}}-2[F_{2}])^{2} \end{equation} 
Hence $ [F_{2}] $ is the acceptable solution of a quadratic equation that is

\begin{equation} [F_{2}]=\left (1+4D[F]_{\textup{tot}} - \sqrt{1+8D[F]_{\textup{tot}}}  \right )/8D \end{equation}

\subsection{Circuit modeling}
The evolution of $ [F]_{\textup{tot}} $ over time depends on the ratio between its synthesis and its elimination. Its rate of synthesis is the maximum transcription frequency "$ s_{\textup{max}} $", weighted by the fraction of time during which the factor is present in the promoter ($ Y $). The loss of $ [F]_{\textup{tot}} $ is simply a linear function of its amount, with a rate constant "$ r $" (removal). Finally, to allow switching to a high expression level, the factor should be synthesised independently of its own auto stimulatory action. This low-frequency synthesis rate is written $ s_{0} $. We thus obtain an ordinary differential equation

\begin{equation} \frac{d[F]_{\textup{tot}}}{dt}=s_{0}+ s_{\textup{max}} \frac{[F_{2}]}{K+[F_{2}]}-r[F]_{\textup{tot}} \end{equation}
that can be expressed as function of $ [F]_{\textup{tot}} $ only, by replacing $ [F_{2}] $ by its value defined in Eq.(5).

\subsection{Draw me a landscape}
A Waddington landscape resembles a potential landscape where basins centered on attractors are separated by barriers of potential. Waddington drew his landscape intuitively without clear mathematical idea (Wang et al., 2011) and several approaches can be proposed to formalize it. In particular, an analogy can be made with the gravitational potential. Gravitational attraction that is inversely proportional to the square of the distance in Newton's law $ F(x)=- M/x^{2} $, can be designed according to Einstein, not as a force applying directly between massive objects, but as a shift to the lowest potential in a curved space described by $ -\int F(x) dx = -M/|x| $, whose representation gives out the popular crater-shaped gravitational attractor. Similarly, the landscape of Waddington can be considered as the negative integral of the gene(s) evolution rate(s) (Ferrell, 2012). In the present case for a unique gene, this integral gives the landscape with two attractors and a separating barrier represented in Fig.2.
This minimalist example is sufficient to show several essential features of gene networks:\\
\noindent
(\textbf{i}) The "spontaneous expression" used for building this landscape can be a transcriptional "noise" related to basal gene promoter (for instance a TATA box).\\
(\textbf{ii}) Nonlinearity is essential to allow such bistability. It is in this case provided by the dimerization of $ F $. Indeed, one can simply verify that in absence of a square in the equations, a unique solution would have been obtained, which means that all cells would have a single gene expression state (monostability). Strongly nonlinear effects of combinations of transcription factors can generate nearly all-or-nothing gene activations and thresholds resembling those of Boolean gene networks whose capacity to give rise to multiple steady states has long been shown (Kauffman, 1969).\\
(\textbf{iii}) The positive feedback loop is also a \textit{sine qua non} condition of multistability (Kaufman et al., 2007). For proof, one can replace the previous loop by a self-inhibition using the approximation of the rapid equilibrium, with a repressor active as a dimer. 
\begin{center}
\includegraphics[width=8cm]{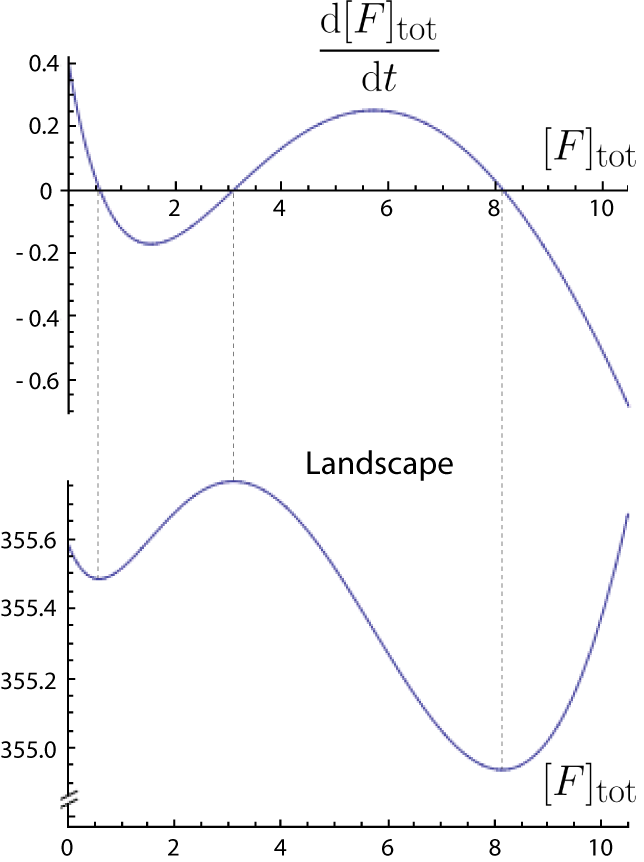} \\
\end{center}
\begin{small} \textbf{Figure 2. Construction of a minimalist unidimensional epigenetic landscape.}\\
Evolution of gene products (top curve) and the corresponding Waddington landscape (bottom curve), as function of the initial condition $ [F]_{\textup{tot}_{0}} $. These profiles have been computed using the following set of parameter values: Basal expression $ s_{0}= 0.4 $; dimerisation constant $ D=0.2 $; DNA dissociation constant $ K=3 $; maximum expression rate $ s_{max} =20 $ and degradation rate $ r=1 $ \end{small}\\
\newline
The simple one-dimensional landscape described here (Fig.2) is purely theoretical because many transcriptional influences interfere in the cell and a gene is rarely controlled by a single factor, but by a combination of factors. Consequently, the so-called production functions that weight the maximum frequency $ s_{\textup{max}} $, include several factors. Depending on the mode of interference between the factors, several production functions can be defined (Bintu et al., 2005). The combined effects of the factors may be cooperative (sigmoidal) or not (hyperbolic) (Michel, 2010), with important consequences for multistability. If the landscape of Waddington has a great pedagogical value in facilitating the visualization of attractors, one should nevertheless beware of false representations. The vertical dimension of the landscape is used to represent the potential, leaving only two dimensions to represent the gene expression levels. A landscape determined by a single gene gives a curve like that of Fig.2 and a landscape determined by two genes appears as a two-dimensional surface deformed in 3D. This image is extensively studied because it is very illuminating to illustrate cell differentiation. In the case of two mutually inhibiting and self-stimulating genes A and B, stem cells are present in an elevated attractor with coexpression of A + B, and two differentiated cell types are found in two low elevation attractors with exclusive expression of either A or B (Huang, 2009; Wang et al., 2011). But landscapes corresponding to three or more genes ($ n $-dimensional) are naturally not representable. As such, the famous Waddington diagram representing a two-dimensional landscape below which a myriad of genes "pull the strings" (Slack, 2002) does not make sense. $ n $-dimensional landscapes are purely mathematical objects, unrepresentable and rarely explicitly computable. But even this mathematical vision is illusory because the thermodynamic and kinetic parameters required for calculation or simulation are almost impossible to determine. The inherently multi-variable nature of biological systems makes them impossible to model in a "bottom-up" manner using elementary kinetic ingredients. Waddington landscapes are essential for their conceptual help in understanding biochemical networks, but are not realistic modeling tools.

\section{How to change attractor}
Multistable networks are not definitely trapped in a single basin of attraction. There are several ways to move from one steady state to another.

\subsection{Passage from a high potential pool to a low potential pool}
This is the most likely evolution, particularly invoked during embryonic development. Stem, totipotent or pluripotent cells are located in elevated basins where the expression of many genes coexist, whereas the differentiated cells are in low potential basins, where the expression of many genes is precluded by incompatible circuits. Cellular switches between attractors are generally multistep processes with a series of binary choices where an upstream valley splits into two  downstream valleys (Foster et al., 2009), thereby offering the theoretical possibility to generate $ 2^{n} $ cell types after $ n $ bifurcations. If transitions accompanied by a drop of potential are predominant, transitions to new states of the same or even of higher potential, are not prohibited. Such jumps have been proposed to challenge the Waddington landscapes (Ladewig et al., 2013), but any transition is just a matter of kinetics and of inputs received from the outside.

\subsection{Passages between basins of the same, or higher elevation}
Lateral jumps between wells of  equivalent potential are involved in trans-differentiation. Jumps from bottom to top attractors underlie de-differentiation, reprogramming of differentiated cells into "induced pluripotent stem cells" (iPSC), and cancer (Huang, 2011). In this respect, primary cancers may occur without genomic alterations (Brock et al., 2009) and might be secondarily consolidated by mutations (Huang and Ingber, 2006). The idea of ​​a reversible cancer without massive oncogenic mutations is not commonly accepted, but it is nevertheless suggested by the possibility of reprogramming the nucleus of malignant cells (Hochedlinger et al., 2004).

\subsection{The procedure for changing attractor}
Genes are the only molecular players in the cell whose number is invariant. Changes of wells result from changes in gene product concentrations and these fluctuations can be programmed or not.

\subsubsection{The road is less important than the destination}
Trajectories between two stationary states are not important by themselves and studying them in an attempt to find so-called molecular relays is not always relevant. An enlightening study showed that the same final state can be reached through different paths (Huang et al., 2005). In this study, the differentiation of HL60 cells into neutrophils can be induced by treatment with either dimethylsulfoxide (DMSO) or retinoic acid. With both treatments, the gene network finally readjusts in the same state; but kinetic transcriptomic studies revealed that intermediate transients are completely different between the two treatments, some genes being upregulated by one treatment and downregulated by the other one. It is therefore clear that a gene induced at an early stage following a perturbation is not necessarily the hallmark of a signalling pathway. In gene networks, transient evolutions do not matter and only the stationary states reflect cellular activities. Accordingly, many experimental cell biologists observed that a new cellular activity can be induced in different ways, which have reciprocal actions. The circular relationships existing in gene networks plunge into deep perplexity biologists who are desperately seeking for originator molecules in order to designate the elusive "therapeutic targets" necessary to obtain medical research funding.

\subsubsection{Unplanned transitions between basins}
The frequency of the jumps out of an energy well depends on the height to climb from the bottom of the well to the top of the surrounding barriers. Evolutionary selected attractors are profound enough to tolerate stochastic fluctuations in gene expression, which are cancelled by interaction circuits. But one cannot exclude that certain combinations of circumstances obtained by chance, in which gene A is slightly upregulated, gene B slightly downregulated etc., concur to destabilize a preexisting attractor. By analogy with statistical mechanics, if we call the total potential difference $ E_{a} $, the output frequency has the form $ k= A \  \textup{exp}(-E_{a}/k_{B}T) $ where $ k_ {B} $ is the Boltzmann constant, $ T $ is the temperature and $ A $ is said to be the pre-exponential factor giving the unit of a rate ($ t^{-1} $). This frequency depends on random molecular fluctuations in the cell, like a wave higher than the others at the surface of the ocean. These events can be forbidden when $ E_{a} $ is high enough, or have an extremely low frequency but may however accidentally occur in a large population of cells.

\subsubsection{Scheduled runs between basins}
To ensure a stereotyped embryonic development with robust cellular states and hierarchical transitions between states, deep wells and acute valleys in the landscape of Waddington, have been selected during evolution. In contrast, the shallower ponds separated by low barriers offer degrees of freedom to the cells, with positive effects on phenotype plasticity but also adverse consequences by allowing the cells to progressively reach an unwanted fatal attractor (Huang, 2011). Exits from energy wells depend on mechanical or chemical signals that affect the transcriptional and post-translational modifications. For example, in the simple case of two attractors in Fig.2, temporary degradation of $ F $ suffices to trigger the switch from the right well to the left one. Once the transition is completed, the degradation can be stopped and the system self-stabilizes in its new attractor. Protein degradation seems to actually play a major role in the reprogramming of eukaryotic cells, as suggested by the large number of genes encoding ubiquitin ligases revealed by genome sequencing. For example, protein phosphorylation triggered by exogenous signals can make the protein a substrate for a preexisting ubiquitin ligases. This general principle can take many other forms. Degradation may be replaced by protein activation or inactivation, for example once again regulated by phosphorylation. The resulting distortion of biochemical network forces the gene network to readjust over a new basin. Through this principle, we see that the cell is in tune with its environment by receiving a wide range of external signals that act on key target proteins capable of destabilizing attractors. These signals include the embryonic inducers of Waddington and Spemann (Slack, 2002) and more generally all chemical or mechanical signals that reprogram all or part of cellular activities. \\
The main lesson of the vision of Waddington is that DNA and genes only set the collection of all possible cellular fates, but that the precise state of a cell is ultimately dictated by the conjunction of chance and of externally imposed conditions. These states can then be maintained by dynamic circuitry only without need for structural marks as long as antagonistic stimuli are not provided.

\section{How to insert chromatin epigenetics in this picture}
\subsection{Chromatin modifications are dispensable for transcriptional memory imprinting}
Cells committed to a particular lineage during embryonic development, maintain their identity over cell divisions. In addition, some cells also have the capacity to memorize exposure to chemicals such as hormones. A famous example is the memory of the estrogen-dependent vitellogenesis by liver cells from egg-laying vertebrates, where the response to estradiol is much faster in animals previously treated with estroadiol in the past. Interestingly if vitellogenin expression is associated with chromatin remodeling of its gene, this remodeling is not persistent enough to explain the memory effect (Burch and Evans, 1986). This example shows that if chromatin remodeling is indeed correlated with the gene expression status, it does not necessarily have a causal influence on its transcriptional memory. Instead, the vitellogenesis memory effect could be purely dynamic (Nicol-Benoit et al., 2011). The examples of bacterial memories, such as that of the lactose operon induction status perpetuated over cellular generations, also proved that life did not await eukaryotic chromatin to make epigenetic memories. Moreover, the origin of this bacterial phenotype maintenance has long been identified as a positive feedback loop (Cohn and Horibata, 1959). "Chromatin epigenetics" are neither more nor less epigenetic than any other enzymatic modification. They contribute to biochemical steady states and thus in sculpting the landscape of Waddington.

\subsection{Chromatin modifications are fleeting executants}
The stationary level of epigenetic "marks" carried by histones results from an incessant ballet of modifications-demodifications, as for any post-translational modification. Accordingly, antagonistic modifying and demodifying enzymes are often present simultaneously in the cells. The fact that chromatin modifications are not set in stone could have been anticipated from old observations such as those of (Thomas et al. 1975), showing that the levels of histone methylation rapidly stabilize at intermediate levels, reflecting specific ratios between methylases and demethylases in the cell. The exchanges of methyl-groups in DNA (CpG methylation) are also very dynamic (Yamagata et al. 2012). The dynamic circuits described previously are recovered for histone modifications for which many positive feedback have been established. The most famous feedback mechanism described in this context is the recruitment of histone-modifying enzymes by neighboring nucleosomes already modified. This allows to strengthen and spread these changes in space along a chromosomal region, and over time through mitosis after dispersion of parental nucleosomes on daughter DNA molecules (Zhu and Reinberg, 2011). This hypothesis has been proposed for the H4K16 deacetylase Sir2 (Imai et al., 2000), the H3K27 methyltransferase of PRC2 (Hansen et al., 2008; Margueron et al., 2009), the SUV39H1/2 H3K9 methyltransferase (Nakayama et al., 2001; Dodd et al. 2007). All these examples of positive loops reconnect chromatin modifications to the more general field of biochemical network circuitry. In addition to transcriptional memory, some very important mechanisms may result from cycles of chromatin modifications-demodifications, such as the ultrasensitivity of program changes (Nicol-Benoit et al., 2012). Permanent modifications-demodification cycles seem unnecessary and wasteful for the cell, and they have therefore been called "futile cycles". But the pioneers of modeling realized that these cycles are far from futile and that this apparent waste has an essential role in cellular decision-making, ultrasensitivity mechanisms such as the "zeroth order" mechanism (Xu and Gunawardena, 2012), operating for one modification (Goldbeter and Koshland, 1981) or cascades of modifications (Ferrell, 1996). The mechanisms established for the cell cycle kinases are naturally valid in the case of histone methylation. These incessant cycles, costly for the cell (Goldbeter and Koshland, 1987), are the price to be paid for a discerning cellular behavior (Michel et al. 2011). It is obvious that the same mechanisms are at work for chromatin modifications (Nicol-Benoit et al., 2012), but this field of investigation has not yet opened and may suffer from a too static view of the famous "epigenetic marks".

\subsection{Technical difficulties}
The most common techniques for studying chromatin modifications are not suitable for measuring the rate of their recycling. The results of chromatin immunoprecipitation assays (ChIP) reflect only stationary states with no information on the dynamics of molecule exchanges at the microscopic level. From this point of view, a common mistake is to assume that changes in chromatin marks observed by ChIP may correspond to microscopic turnovers of these marks, which is false. For example, a delayed negative feedback effect usually causes large oscillations of the considered biological activity (in the manner of a slow thermostat), but the periods of these cycles are completely unrelated to the underlying molecular turnovers. To get an idea of the actual frequency of microscopic turnovers, we must turn to more complicated techniques such as pulse-chase experiments. Such approaches have been developed to study the replacement of nucleosomes (Deal and Henikoff, 2010), DNA methylation (Yamagata et al. 2012) and histone modifications (Thomas et al., 1975; Ng et al., 2009; Zee et al. 2010). The latter, which require complex techniques, give renewal times in the range of day(s), unable to ensure static epigenetic memory for years.

\subsection{Chromatin modifications: receivers or decision makers?}
A specificity of the network concept is that the functions of decision and execution are not separable. In essence, a network operates democratically and each node in the network is just a cog. This is obviously the case for chromatin modifications that are both the results and regulators of gene expression, taking place in circular interrelations. As opposed to a decision-making role of the cellular phenotype "above the genome" at the tip of the pyramid, chromatin could sometimes play a simple role of receptor of phenotype changes imposed exogenously. A recent study showed that a mechanically induced cell stretching could both initiate actin polymerization and reprogram gene expression by decompaction of chromatin (Iyer et al., 2012). We knew that the expression of specific genes can cause a reorganization of the cytoskeleton, for example through induction of Serum Responsive Factor (SRF)-responsive genes. Now we see that the reverse is also true. An other role of chromatin modifications in signal integration is emerging in the field of cellular metabolism. Given that most substrate precursors of enzymatic chromatin modifications are byproducts of metabolism (including NADH, acetyl-CoA or SAM), their level can dictate the extent of chromatin modifications (Kim et al., 2005, Wallace and Fan, 2010; Sassone-Corsi, 2013). \\
\newline
In conclusion, to date only the sequence of nucleotides in DNA can be reasonably regarded as a structural memory. All others, including chromatin marks, belong to the dynamic epigenetic landscape of Waddington. Waddington's vision is not obsolete, as sometimes heard in the introduction of lectures on chromatin, but it is instead the general framework in which chromatin should be inserted.

\section{References}
\begin{small}
Bintu, L., Buchler, N.E., Garcia, H.G., Gerland, U., Hwa, T., Kondev, J. and Phillips, R. (2005) Transcriptional regulation by the numbers: models. Curr. Opin. Genet. Dev. \textbf{15}, 116-124\\
Brock, A., Chang, H. and Huang, S. (2009) Non-genetic heterogeneity--a mutation-independent driving force for the somatic evolution of tumours. Nat. Rev. Genet. \textbf{10}, 336-342\\
Burch, J.B. and Evans, M.I. (1986) Chromatin structural transitions and the phenomenon of vitellogenin gene memory in chickens. Mol. Cell. Biol. \textbf{6}, 1886-1893\\
Cohn, M. and Horibata, K. (1959) Inhibition by glucose of the induced synthesis of the beta-galactoside-enzyme system of \textit{Escherichia coli}. Analysis of maintenance. J. Bacteriol. \textbf{78}, 601-612\\
Deal, R.B. and Henikoff, S. (2010) Catching a glimpse of nucleosome dynamics. Cell Cycle \textbf{9}, 3389-3390\\
Dodd, I.B., Micheelsen, M.A., Sneppen, K. and Thon, G. (2007) Theoretical analysis of epigenetic cell memory by nucleosome modification. Cell \textbf{129}, 813-822\\
Ferrell, J.E.Jr. (1996) Tripping the switch fantastic: how a protein kinase cascade can convert graded inputs into switch-like outputs. Trends Biochem. Sci. \textbf{21}, 460-466\\
Ferrell, J.E.Jr. (2012) Bistability, bifurcations, and Waddingtons epigenetic landscape. Curr. Biol. \textbf{22}, R458-R466\\
Foster, D.V., Foster, J.G., Huang, S. and Kauffman, SA. (2009) A model of sequential branching in hierarchical cell fate determination. J. Theor. Biol. \textbf{260}, 589-597\\
Goldbeter, A. and Koshland, D.E.Jr. (1981) An amplified sensitivity arising from covalent modification in biological systems. Proc. Natl. Acad. Sci. USA \textbf{78}, 6840-6844\\
Goldbeter, A. and Koshland, D.E.Jr. (1987) Energy expenditure in the control of biochemical systems by covalent modification. J. Biol. Chem. \textbf{262}, 4460-4471\\
Hansen, K.H., Bracken, A.P., Pasini, D., Dietrich, N., Gehani, S.S., Monrad, A., Rappsilber, J., Lerdrup, M. and Helin, K. 2008. A model for transmission of the H3K27me3 epigenetic mark. Nat Cell Biol. \textbf{10}, 1291-1300\\ 
Hochedlinger, K., Blelloch, R., Brennan, C., Yamada, Y., Kim, M., Chin, L. and Jaenisch, R. (2004) Reprogramming of a melanoma genome by nuclear transplantation. Genes Dev. \textbf{18}, 1875-1885\\
Huang, S. Eichler, G., Bar-Yam, Y. and Ingber, D.E. (2005) Cell fates as high-dimensional attractor states of a complex gene regulatory network. Phys. Rev. Let. \textbf{94}, 128701\\
Huang, S. and Ingber, D.E. (2006) A non-genetic basis for cancer progression and metastasis: self-organizing attractors in cell regulatory networks. Breast Dis. \textbf{26}, 27-54\\
Huang, S. (2009) Reprogramming cell fates: reconciling rarity with robustness. Bioessays \textbf{31}, 546-560\\ 
Huang, S. (2011) On the intrinsic inevitability of cancer: from foetal to fatal attraction. Semin. Cancer Biol. \textbf{21}, 183-199\\
Imai, S., Armstrong, C.M., Kaeberlein, M. and Guarente, L. (2000) Transcriptional silencing and longevity protein Sir2 is an NAD-dependent histone deacetylase. Nature \textbf{403}, 795-800\\
Iyer, K.V., Pulford, S., Mogilner, A. and Shivashankar, G.V. (2012) Mechanical activation of cells induces chromatin remodeling preceding MKL nuclear transport. Biophys J. \textbf{103}, 1416-1428\\
Kauffman, S.A. (1969) Metabolic stability and epigenesis in randomly constructed genetic nets. J. Theor. Biol. \textbf{22}, 437-467\\
Kaufman, M, Soul\'e, C. and Thomas, R. (2007) A new necessary condition on interaction graphs for multistationarity. J. Theor. Biol. \textbf{248}, 675-685\\
Keller, A.D. (1995) Model genetic circuits encoding autoregulatory transcription factors. J. Theor. Biol. \textbf{172}, 169-185\\
Kim, J.H., Cho, E.J., Kim, S.T. and Youn, H.D. (2005) CtBP represses p300-mediated transcriptional activation by direct association with its bromodomain. Nat. Struct. Mol. Biol. \textbf{12}, 423-428\\
Ladewig, J., Koch, P. and Br\"ustle, O. (2013) Leveling Waddington: the emergence of direct programming and the loss of cell fate hierarchies. Nat. Rev. Mol. Cell. Biol. \textbf{14}, 225-236\\ 
Slack, J.M.W. (2002) Conrad Hal Waddington: the last Renaissance biologist? Nat. Rev. Genet. \textbf{3}, 889-895\\
Thomas, G., Lange, H.W. and Hempel, K. (1975) Kinetics of histone methylation in vivo and its relation to the cell cycle in Ehrlich ascites tumor cells. Eur. J. Biochem. \textbf{51}, 609-615\\
Margueron, R., Justin, N., Ohno, K., Sharpe, M.L., Son, J., Drury, W.J. 3rd, Voigt, P., Martin, S.R., Taylor, W.R., De Marco, V., Pirrotta, V., Reinberg, D. and Gamblin, S.J. (2009) Role of the polycomb protein EED in the propagation of repressive histone marks. Nature \textbf{461}, 762-767\\
Michel, D. (2009) Fine tuning gene expression through short DNA-protein binding cycles. Biochimie \textbf{91}, 933-941\\
Michel, D. (2010) How transcription factors can adjust the gene expression floodgates. Prog. Biophys. Mol. Biol. \textbf{102}, 16-37\\
Michel D. (2011) Basic statistical recipes for the emergence of biochemical discernment. Prog. Biophys. Mol. Biol. \textbf{106}, 498-516\\
Michel, D. (2013) Life is a self-organizing machine driven by the informational cycle of Brillouin. Orig. Life Evol. Biosph. \textbf{43}, 137-150\\
Nakayama, J., Rice, J.C., Strahl, B.D., Allis, C.D. and Grewal, S.I. (2001) Role of histone H3 lysine 9 methylation in epigenetic control of heterochromatin assembly. Science \textbf{292}, 110-113\\
Ng, S.S., Yue, W.W., Oppermann, U. and Klose, R.J. (2009) Dynamic protein methylation in chromatin biology. Cell. Mol. Life Sci. \textbf{66}, 407-422\\
Nicol-Benoit, F. Amon, A. Vaillant, C. Le Goff, P., Le Dréan, Y. Pakdel, F. Flouriot, G. Valotaire, Y. and Michel, D. (2011) A dynamic model of transcriptional imprinting derived from the vitellogenesis memory effect. Biophys. J. \textbf{101}, 1557-1568\\
Nicol-Benoit, F., Le-Goff, P., Le-Dréan, Y., Demay, F., Pakdel, F., Flouriot, G. and Michel, D. (2012) Epigenetic memories: Structural marks or active circuits? Cell. Mol. Life Sci. \textbf{69}, 2189-2203\\ 
Sassone-Corsi, P. (2013) When metabolism and epigenetics converge. Science \textbf{339}, 148-150\\
Wallace, D.C. and Fan, W. (2010) Energetics, epigenetics, mitochondrial genetics. Mitochondrion \textbf{10}, 12-31\\
Wang, J. Zhang, K. Xu, L. and Wang, E. (2011) Quantifying the Waddington landscape and biological paths for development and differentiation. Proc. Natl. Acad. Sci. USA \textbf{108}, 8257-8262\\
Xu, Y. and Gunawardena, J. (2012) Realistic enzymology for post-translational modification: zero-order ultrasensitivity revisited. J. Theor. Biol. \textbf{311}, 139-152\\
Yamagata, Y., Szab\'o, P., Sz\"uts, D., Bacquet, C., Ar\`anyi, T. and P\'aldi, A. (2012)
Rapid turnover of DNA methylation in human cells. Epigenetics \textbf{7}, 141-145\\
Zee, B.M., Levin, R.S., Xu, B., LeRoy, G., Wingreen, N.S. and Garcia, B.A. (2010) In vivo residue-specific histone methylation dynamics. J. Biol. Chem. \textbf{285}, 3341-3350\\
Zhu, B. and Reinberg, D. (2011) Epigenetic inheritance: uncontested? Cell Res. \textbf{21}, 435-441
\end{small}
\end{document}